\documentclass[referee]{mn2e}
\usepackage{epsfig}
\usepackage{txfonts}

\title[The role of $r$-mode damping in the thermal evolution of neutron stars]{ The role of $r$-mode damping in the thermal evolution of neutron stars}

\author[Shu-Hua Yang, Xiao-Ping Zheng, Chun-Mei Pi, Yun-Wei Yu]
{Shu-Hua Yang $^{}$\thanks{E-mail: ysh@phy.ccnu.edu.cn}, Xiao-Ping Zheng
$^{}$\thanks{E-mail: zhxp@phy.ccnu.edu.cn}, Chun-Mei Pi, Yun-Wei Yu
\\ Institute of Astrophysics, Huazhong Normal
University, Wuhan 430079, China}

\date{Accepted 0000, Received 0000}
\pagerange{\pageref{firstpage}--\pageref{lastpage}} \pubyear{0000}

\begin{document}
\label{firstpage} \maketitle

\begin{abstract}

The thermal evolution of neutron stars (NSs) is investigated by coupling with the evolution of $\textit{r}$-mode instability that is described by a second order model.
The heating effect due to shear viscous damping of the $\textit{r}$-modes enables us to understand the high temperature of two young pulsars (i.e., PSR B0531+21 and RX J0822-4300) in the framework of the simple $npe$ NS model, without superfluidity or exotic particles.
Moreover, the light curves predicted by the model within an acceptable parameter regime may probably cover all of the young and middle-aged pulsars in the $\lg T_s^{\infty}-\lg t$ panel, and an artificially strong $p$ superfluidity invoked in some early works is not needed here. Additionally, by considering the radiative viscous damping of the $\textit{r}$-modes, a surprising extra cooling effect is found, which can even exceed the heating effect sometimes although plays an ignorable role in the thermal history.

\end{abstract}


\section{Introduction}

The composition of neutron star (NS) interior is still poorly known due to the
uncertainties of nuclear physics. It has been hoped that comparing theoretical thermal
evolution of NSs with thermal emission data from observations could yield information
about their internal properties. Our knowledge of the cooling history
of a NS has been improving as we were refining the physical ingredients that
play a key role on the thermal evolution of NSs.
Page et al. (2004) proposed the minimal cooling model and
suggested that four basic physical inputs should be considered
in the simulation of NS thermal evolution, namely,  the equation of state,
superfluid properties of relevant components, envelope composition and stellar mass.
Aguilera et al. (2008) argued that the magnetic field is another
basic physical ingredient should not be ignored.  In fact,
heating mechanisms should not be ignored, too.

During the evolution of NSs, heating mechanisms may be
present and dramatically change the thermal evolution of the star.
Several heating mechanisms have been extensively discussed, for
example, rotochemical heating (Reisenegger 1995; Fern\'andez \& Reisenegger 2005),
vortex creep heating (Umeda et al. 1995), Joule heating (Aguilera et al. 2008),
heating due to the hardron-quark transition (Kang \& Zheng 2007) and heating
due to \textit{r}-mode damping (Zheng et al. 2006; Yu et al. 2009).

\textit{R}-modes in a perfect fluid star with arbitrary rotation
arise due to the action of the Coriolis force with positive
feedback (Andersson 1998; Friedman \& Morsink 1998), succumbing to
gravitational radiation-driven Chandrasekhar-Friedman-Schutz
instability. However, in a realistic star, the \textit{r}-mode
evolution is determined by the competition between
damping effect due to viscous dissipation
and the destabilizing effect due to gravitational
radiation. Based on the conservation of angular momentum, a
phenomenological model describing the \textit{r}-mode evolution
was proposed by Owen et al. (1998) and improved by Ho \& Lai
(2000).  In this original version of the model, an unbounded growth could lead the
modes to an unphysical regime because nonlinear effects are ignored.
As an important nonlinear effect, differential rotation induced by
\textit{r}-modes was first studied by Rezzolla et al. (2000, 2001)
and confirmed by some numerical studies (Stergioulas \& Font 2001; Lindblom et al. 2001).
S\'a (2004) solved the fluid equations within nonlinear theory up to the
second order in the mode amplitude and described the differential
rotation analytically. By extending the \textit{r}-mode evolution
model of Owen et al. (1998) to this nonlinear case, S\'a \& Tom\'e
(2005, 2006) obtained a saturation amplitude of \textit{r}-modes
self-consistently. Yu et al. (2009) studied the long-term spin and thermal
evolution of isolated NSs under the influence of the differential rotation.
They found that the stars can keep nearly a constant temperature for over a thousand years
since the differential rotation can significantly prolong the
duration of \textit{r}-modes.

In this paper, we study the thermal evolution of isolated NSs
with differential rotation induced by \textit{r}-modes using a realistic
equation of state (EOS). The same EOS is used by Kaminker et al. (2001)
in the simulation of NS cooling, but their interpretation of the observations requires
a strong $p$ superfluidity. As pointed by Tsuruta et al. (2002) and Blaschke et al. (2004),
the strong $p$ superfluidity contradict the microphysical calculations
in case of high enough proton concentration to permit nucleon direct Urca process.
The inclusion of \textit{r}-mode dissipation would probably resolve this contradiction.
In this work, we also take into account the recently
realized radiative viscosity (Sa'd et al. 2009).

After the introduction in this section, Section 2 presents the
formulism of \textit{r}-mode evolution and the spin evolution
of NSs up to second order in the mode amplitude, Section 3
displays the equation of thermal evolution and the heating (or cooling) term
due to viscous dissipation of \textit{r}-modes. The physical inputs and
our results are given in Section 4, and the conclusions
and discussions are presented in the last section.

\section{The {\it r}-mode evolution and the spin evolution of NSs to second-order  }
For a rotating barotropic Newtonian star, the \textit{r}-mode
solutions of perturbed fluid equations can be found in spherical
coordinates ($r, ~\theta, ~\phi$) at first order in $\alpha$ as
(Lindblom et al. 1998),
\begin{eqnarray}
\delta^{(1)}v^{r}&=&0,\\
\delta^{(1)}v^{\theta}&=&\alpha\Omega C_ll\left({r\over
R}\right)^{l-1}\sin^{l-1}\theta\sin(l\phi+\omega
t),\\
\delta^{(1)}v^{\phi}&=&\alpha\Omega C_ll\left({r\over
R}\right)^{l-1}\sin^{l-2}\theta\cos\theta\cos(l\phi+\omega t),
\end{eqnarray}
and at second order in $\alpha$ as (S\'a 2004)
\begin{eqnarray}
\delta^{(2)}v^{r}&=&\delta^{(2)}v^{\theta}=0,\\
\delta^{(2)}v^{\phi}&=&{1\over2}\alpha^2\Omega
C_l^2l^2(l^2-1)\left({r\over
R}\right)^{2l-2}\sin^{2l-4}\theta\nonumber\\
&&+\alpha^2\Omega Ar^{N-1}\sin^{N-1}\theta,
\end{eqnarray}
where $\alpha$ represents the amplitude of the oscillation, $R$
and $\Omega$ are the radius and angular velocity of the
unperturbed star, $\omega=-\Omega(l+2)(l-1)/(l+1)$,
$C_l=(2l-1)!!\sqrt{(2l+1)/[2\pi(2l)!l(l+1)]}$, $A$ and $N$ are two
constants determined by the initial condition. For simplicity,
S\'a \& Tom\'e (2005) suggested $N=2l-1$ and redefined $A$ by
introducing a new free parameter $K$ as
$A={1\over2}KC_l^2l^2(l+1)R^{2-2l}$.
For the most unstable $l=2$ \textit{r}-mode of primary interest to
us, the second-order solution $\delta^{(2)}v^{\phi}$ shows a
differential rotation of the star induced by the \textit{r}-mode
oscillation, i.e., large scale drifts of fluid elements along
stellar latitudes.

Using $\delta^{(1)}v^{i}$ and
$\delta^{(2)}v^{i}$, the corresponding Lagrangian displacements
$\xi^{(1)i}$ and $\xi^{(2)i}$ can be derived and then the physical
angular momentum of the $l=2$ \textit{r}-mode can be calculated up
to the second order in $\alpha$ as (S\'a 2004; S\'a \& Tom\'e
2005)
\begin{eqnarray}
J_r=J^{(1)}+J^{(2)}={{(4K+5)}\over2}\alpha^2\tilde{J}MR^2\Omega,
\end{eqnarray}
where $\tilde{J}=1.635\times10^{-2}$ and
\begin{eqnarray}
J^{(1)}&=&-\int\rho\partial_{\phi}\xi^{(1)i}\left(\partial_{t}\xi^{(1)}_i+v^{k}\nabla_{k}\xi^{(1)}_i\right)dV,\\
J^{(2)}&=&{1\over\Omega}\int\rho
v^{i}\left[\partial_t\xi^{(1)k}\nabla_i\xi^{(1)}_k+v^{k}\nabla_k\xi^{(1)m}\nabla_i\xi^{(1)}_m
+\partial_t\xi^{(2)}_i\right.\nonumber\\
&&\left.+v^{k}\left(\nabla_i\xi^{(2)}_k+\nabla_k\xi^{(2)}_i\right)\right]dV.
\end{eqnarray}
Meanwhile, following Owen et al. (1998) and S\'a (2004), we
further express the energy of the $l=2$ \textit{r}-mode by
\begin{eqnarray}
E_r=J^{(2)}\Omega-{1\over3}J^{(1)}\Omega={(4K+9)\over2}\alpha^{2}\tilde{J}MR^{2}\Omega^{2}.
\end{eqnarray}
When $K=-2$, $J^{(2)}$ vanishes and the expressions of $J_r$ and
$E_r$ return to their canonical forms (Owen et al. 1998), in other
words, the differential rotation disappears.
Both the physical
angular momentum and energy of \textit{r}-modes are increased by
gravitational radiation back reaction and decreased by viscous
damping, which yields
\begin{eqnarray}
{dJ_r\over dt}&=& {2J_r\over\tau_g}-{2J_r\over\tau_v},\label{jrt}\\
{dE_r\over dt}&=& {2E_r\over\tau_g}-{2E_r\over\tau_v},\label{ert}
\end{eqnarray}
where $\tau_g$ is the growth timescale due to gravitational-wave emission,  $\tau_{v}=(\tau_{sv}^{-1}+\tau_{bv}^{-1}+\tau_{rv}^{-1})^{-1}$ is the damping
timescale due to viscous dissipation; $\tau_{sv}$,$\tau_{bv}$ and $\tau_{rv}$ are
the timescales of the shear viscous damping, bulk viscous damping and
radiative viscous damping (the radiative viscous will interpreted latter), respectively.

From equation ($\ref{jrt}$) and ($\ref{ert}$), it can be seen that
the \textit{r}-modes are unstable if $(\tau_g^{-1}-\tau_v^{-1})^{-1}>0$.
In this case, a small perturbation would lead to a non-ignorable growth of the modes.
The competition between the gravitational destabilizing effect
that is dependent on $\Omega$ and the $T$-dependent viscous
damping effect determines an instability window in the $T-\Omega$
plane.

For a normal NS with a strong magnetic field
($\sim10^{10-12}$ G), besides the braking effect due to
gravitational radiation, the spindown of the star resulting from
magnetic dipole radiation should also be taken into account. So,
we ought to write the decrease of the total angular momentum of
the star as (Owen et al. 1998; Ho \& Lai 2000; S\'a \& Tom\'e
2005)
\begin{equation}
{dJ\over dt}=
-{3\alpha^2\tilde{J}MR^{2}\Omega\over\tau_g}-{I\Omega\over\tau_m}\label{jt},
\end{equation}
where $\tau_m=1.35\times10^{9}B_{12}^{-2}(\Omega/\sqrt{\pi
G\bar{\rho}})^{-2}$s is the magnetic braking timescale and
$I=\tilde{I}MR^2$ with $\tilde{I}=0.261$ is the moment of inertial
of the star. Due to the \textit{r}-mode oscillation, the total
angular momentum of the star could be separated into two parts,
i.e., $J=I\Omega+J_r$. Then, Eqs. (\ref{jrt}) and (\ref{jt}) yield
\begin{eqnarray}
{d\alpha\over
dt}&=&\left[1+{4\over3}(K+2)Q\alpha^{2}\right]{\alpha\over\tau_g}-\left[1+{1\over3}(4K+5)Q\alpha^{2}\right]
{\alpha\over\tau_v}+{\alpha\over2\tau_m}\label{alphat} ,\\
{d\Omega\over dt}&=&-{8\over3}(K+2)Q\alpha^{2}{\Omega\over\tau_g}
+{2\over3}(4K+5)Q\alpha^{2}{\Omega\over\tau_v}-{\Omega\over\tau_m},\label{omegat}
\end{eqnarray}
where $Q=3\tilde{J}/2\tilde{I}=0.094$.

\section{Thermal Evolution of NSs}

The equation of thermal evolution can be written as (Yakovlev et al. 1999; Yakovlev \& Pethick 2004)
\begin{equation}
C_{V}\frac{dT}{dt} = -L_{\nu} -L_{\gamma} +H_{v}-\Delta L_{\nu},\label{tempt}
\end{equation}
where $C_{V}$ is the total stellar heat capacity. In NSs composed of simple $npe$ matter, the electrons
constitute an almost ideal, strongly degenerate, ultra-relativistic gas; neutrons and protons constitute
a non-relativistic strong non-ideal Fermi liquid.

Meanwhile, $L_{\nu}$ is the luminosity of neutrinos
generated in numerous reactions in the interiors of neutron stars.
The main processes we used are nucleon direct
Urca, nucleon modified Urca and nucleon bremsstrahlung.
If the proton and electron Fermi momenta are
too small compared with neutron Fermi momenta, the nucleon direct
Urca process is forbidden because it is impossible to satisfy conservation of
momentum (Lattimer et al. 1991). Under typical conditions,
 one finds that the ratio of the
number density of protons to that of nucleons must exceed about
0.11 for the process to be allowed.

$L_{\gamma}$ is the surface photon luminosity given by
\begin{equation}
L_{\gamma}=4\pi R^{2}\sigma T_{s}^{4},
\end{equation}
here $\sigma$ is the Stefan-Boltzmann constant and $T_{s}$ is the
surface temperature.The relation between $T_{s}$ and the internal NS temperature $T$ is taken from
Potekhin et al. (1997), supposed the outer heat blanketing NS envelope
is made of ion and neglecting the effects of surface magnetic fields.
We also note that the effective surface temperature detected by a distant observer is $T_{s}^{\infty}=T_{s}\sqrt{1-R_{g}/R}$, where $R_{g}$ is the gravitational stellar radius.

($H_{v}-\Delta L_{\nu}$) is the energy per unit time induced by viscous dissipation of \textit{r}-modes.
As we can see from equation ($\ref{ert}$), one part of the oscillation energy of \textit{r}-modes
is converted into heat energy through shear viscous damping
and  bulk viscous damping($H_{v}=2E_r (\tau_{sv}^{-1} + \tau_{bv}^{-1})$), and the other part
is converted into neutrino emissivity through radiative
viscous damping($\Delta L_{\nu}=2E_r \tau_{rv}^{-1}$). As a result
\begin{eqnarray}
H_{v}-\Delta L_{\nu}=2E_r ( \frac{1}{\tau_{sv}} + \frac{1}{\tau_{bv}} -  \frac{1}{\tau_{rv}}) ,
\end{eqnarray}
using equation (\ref{taurv}) below, we get
\begin{eqnarray}
H_{v}-\Delta L_{\nu}=2E_r ( \frac{1}{\tau_{sv}} - \frac{1}{2}\frac{1}{\tau_{bv}} ) .
\end{eqnarray}
Hence, if $\tau_{sv} < 2\tau_{bv}$, we have $(H_{v}-\Delta L_{\nu})>0$, the star would
be heated by viscous dissipation of \textit{r}-modes ;
and if $\tau_{sv}>2\tau_{bv}$, we have $(H_{v}-\Delta L_{\nu})<0$,
viscous dissipation causes an extra cooling of the star.

\section{Physical inputs and results}

\subsection{NS EOS}

In the simulation of thermal evolution, we employ
the simplest possible nuclear composition, namely neutrons ($n$),
protons ($p$) and electrons ($e$). We adopt a moderately stiff equation of state (EOS) of this
matter proposed by Prakash et al. (1988) (their model I with the compression modulus of
saturated nuclear matter $K=240$MeV). The maximum mass of this model is $M=1.977M_{\odot}$,
and the direct Urca process is forbidden at $M<M_{D}=1.358M_{\odot}$.

\subsection{The $r$-mode timescales}
The calculation of $r$-mode  timescales are very complicated,
and until now there are no calculations based on realistic EOS.
The following timescales (for $l=2$ $r$-modes) we employed in
this paper are obtained with a polytropic
equation of state as $p=k\rho^2$ for NSs, with $k$ chosen so that
the mass and radius of the star are $M=1.4M_{\odot}$ and $R=12.53$
km. In the following equations, the conventions $T_9\equiv T/10^9$ and
$\tilde{\Omega}\equiv\Omega/\sqrt{\pi G\bar{\rho}}$ are used.

The gravitational radiation timescale is (Owen et al. 1998)
\begin{equation}
\tau_g=3.26\tilde{\Omega}^{-6}s \label{taug}.
\end{equation}

The shear viscous damping timescale (due to the neutron-neutron scattering) is (Owen et al. 1998)
\begin{equation}
\tau_{sv}=2.52\times10^{8}T_9^{2}s.
\end{equation}

In the case of $npe$ matter where only the modified Urca process is relevant,
the bulk viscous damping timescale is (Andersson \& Kokkotas 2001)
\begin{equation}
\tau_{bv}(MUrca)=1.20\times10^{11}T_9^{-6}\tilde{\Omega}^{-2}s,
\end{equation}
and in the case of direct Urca process (Owen et al. 1998)
\begin{equation}
\tau_{bv}(DUrca)=6.99\times10^{8}T_9^{-6}\tilde{\Omega}^{-2}s.
\end{equation}

Recently, Sa'd et al. (2009) first demonstrated that there exists a new mechanism for
damping the energy of stellar oscillations, namely the radiative viscous dissipation.
Urca processes contribute to  the damping of density
perturbations not only by converting energy into heat via bulk viscosity,
but also by converting it into an increase of the neutrino emissivity via
radiative viscosity. They found the radiative viscosity coefficient is 1.5 times
larger than the bulk viscosity coefficient. Thus, the damping time scale of
radiative viscosity is
\begin{equation}
\tau_{rv}=\frac{2}{3} \tau_{bv}.\label{taurv}
\end{equation}

Since $\tau_{bv}(MUrca)$ is about three orders larger than $\tau_{bv}(DUrca)$,
it is natural that for the two kinds of NSs the direct Urca process
is permitted or not, the \textit{r}-mode instability windows
(a window determined by  $\tau_g^{-1}-\tau_v^{-1}=0$ in the $T-\Omega$
plane) are quite different.
In contrast, the time evolution behavior of $\alpha$ and $\Omega$
for these two different kinds of NSs are similar.
We can see this from eqs.(\ref{alphat}) and (\ref{omegat})
(the following argument is based on the relation $\tau_g << \tau_m$,
which is correct unless the magnetic fields of NS is extremely large).
For a nascent NS with $T\sim10^{10}$K and $\Omega\sim\frac{2}{3}\sqrt{\pi G \bar{\rho}}$,
we can easily find $\tau_g<<\tau_{sv}$ and $\tau_g<<\tau_{bv}$.
This means the terms of $\tau_g$
in eqs. (\ref{alphat}) and (\ref{omegat}) are decisive in the
early evolution of NS for both cases $\tau_{bv}=\tau_{bv}(DUrca)$
and $\tau_{bv}=\tau_{bv}(MUrca)$.
More exactly, fig.3 in Yu et al. (2009) (they adopted $\tau_{bv}=\tau_{bv}(DUrca)$)
indicates that not only
for a nascent NS but also for NS of phase I, II and III,
$\tau_g$ dominates. Meanwhile, during the following
phase IV and V in their figure, $\tau_{sv}<<\tau_{bv}$, $\tau_g$ and $\tau_{sv}$
control the evolution of NSs. Thus, we conclude that $\tau_{bv}$
plays an ignorable role during the evolution of NSs.



\subsection{The results}
We calculated Equations (\ref{alphat}),  (\ref{omegat}) and (\ref{tempt}) numerically,
taking the initial temperature $T_0=10^{10}$K, the
initial \textit{r}-mode amplitude $\alpha_0=10^{-6}$, the
initial angular velocity $\Omega_0=\frac{2}{3}\sqrt{\pi G \bar{\rho}}$ and
the magnetic field $B=10^{12}$ G.

Fig.1 shows the evolution of $\alpha$, $\Omega$ and $T$ of a
$1.4M_{\odot}$ NS and $K=1000$.
We can see from  Fig.1 (c) that the NS core
can keep high temperature for $4.6\times10^{3}$(or $10^{3.66}$)years.
As illustrated by Yu et al. (2009), during the early part of the \textit{r}-mode
evolution, the rotation energy of the star (${1\over2}I\Omega^2$) is
converted into the oscillation energy,
the internal energy, and the energy of gravitational waves.
Nevertheless, during the late part, the energy
deposited in the \textit{r}-modes would be released gradually via
heating the star and accelerating the stellar rotation.

Fig.2 (the parameters are the same as Fig.1) displays
the evolution curves of $(H_{v}-\Delta L_{\nu})$
due to viscous dissipation and ($-L_{\nu} -L_{\gamma}$).
The evolution of $(H_{v}-\Delta L_{\nu})$ can be divided into
three stages: (1) $lgt<-4.61$. In this phase,
the relation $\tau_{sv}>2\tau_{bv}$ is fulfilled,
and the viscous dissipation of \textit{r}-modes results in
an extra cooling to the thermal evolution of NS.
This extra cooling never plays an important role during the evolution of the NS
because it presents in the early part of NS evolution
where the  star interior is still very hot, and it is too small
comparing with the neutrino luminosity $L_{\nu}$.
Moreover, it's easy to understand that
the same results can be reached to NS
of small mass ($M<1.358M_{\odot}$) where the direct Urca process is forbidden.
(2) $-4.61<lgt<3.65$. In this phase, $\tau_{sv}<2\tau_{bv}$,
the \textit{r}-mode energy is dissipated mainly by
shear viscosity and the star is heated.
(3) $lgt>3.65$. In this phase, $(H_{v}-\Delta L_{\nu})=0$, since the \textit{r}-modes disappear
in this stage (see Fig.1(a)).

 Fig.3 shows the surface temperature $T_{s}^{\infty}$ evolution of
 $M=1.3M_{\odot}$, $M=1.365M_{\odot}$ and $M=1.4M_{\odot}$ NS with different $K$.
 Note that the direct Urca process is forbidden at
 $M<M_{D}=1.358M_{\odot}$ in our EOS model. We can see that
 all curves taking into account \textit{r}-mode dissipation can explain
 two young and hot pulsar data (PSR B0531+21 and RX J0822-4300)
 when taking proper value of $K$ ($K=100$ for $1.3M_{\odot}$ and $M=1.365M_{\odot}$
, $K=1000$ for $1.4M_{\odot}$ ).
 Maybe the $1.3M_{\odot}$ ($K=100$) curve looks too high
 for the explanation of the two data. While, it should be emphasized we don't
 expect our model to explain the observation data precisely; because this model still have
 some uncertainties, such as the realistic non-linear evolution
 of \textit{r}-mode instability is not clear and the timescales
 we used are not based on the specified EOS. Nonetheless, it is
 certain that the dissipation of the oscillations can
 provide large enough heat to raise the temperature of young neutron stars,
 and this isn't a negligible effect.

\section*{5. Conclusions and discussions}
We have studied the thermal evolution of NSs, considering the influence
of \textit{r}-mode instability to second order.
For the first time we take into account the radiative viscosity, and
find the radiative viscous dissipation of \textit{r}-modes results in an extra cooling of
NSs; while in former studies it is taken for granted
that viscous damping would lead to the heating of stars. However,
we find that this extra cooling can be well neglected in the thermal evolution history of NSs.

On the other hand, the NS is heated due to shear viscous damping of \textit{r}-modes,
and it can keep a high temperature for several thousand years, even tens of thousands of years.
This enables us to explain two young and hot pulsar data (PSR B0531+21 and RX J0822-4300)
with NS model composed of only $npe$ matter, without superfluidity or exotic particles.
In contrast, under the same NS EOS, Kaminker et al. (2001) explained these data
by the inclusion of strong $p$ superfluidity (the maximum critical temperature $T_{cp}\gtrsim 5\times 10^{9}$K).
The superfluidity they employed is too strong and it was doubted by
many works (Tsuruta et al. 2002; Blaschke et al. 2004); because
Takatsuka \& Tamagaki (1997) showed, through careful microphysical
calculations, that for neutron matter with such high proton concentration as
to permit nucleon direct Urca process, the superfluid critical temperature should be extremely
low, about several $\times 10^{7}$K.
However, if consider a wider value range of NS mass and K, our light curves
may probably cover all of the young and middle-aged thermal emission data,
and the artificially strong $p$ superfluidity invoked in Kaminker et al. (2001) is no longer needed .
Therefore, the explanation of observation data maybe doesn't contradict
the calculation of microphysics.  Of course, we don't expect our
result to fit the observation data accurately, but future
studies using an improved model, or even including other heating effect would
further improve our results.

\section*{Acknowledgments}
 This work is supported by NFSC under Grant Nos.10773004 and 10603002.

\clearpage

 \begin{figure}
   \centering
   \includegraphics[width=0.8\textwidth]{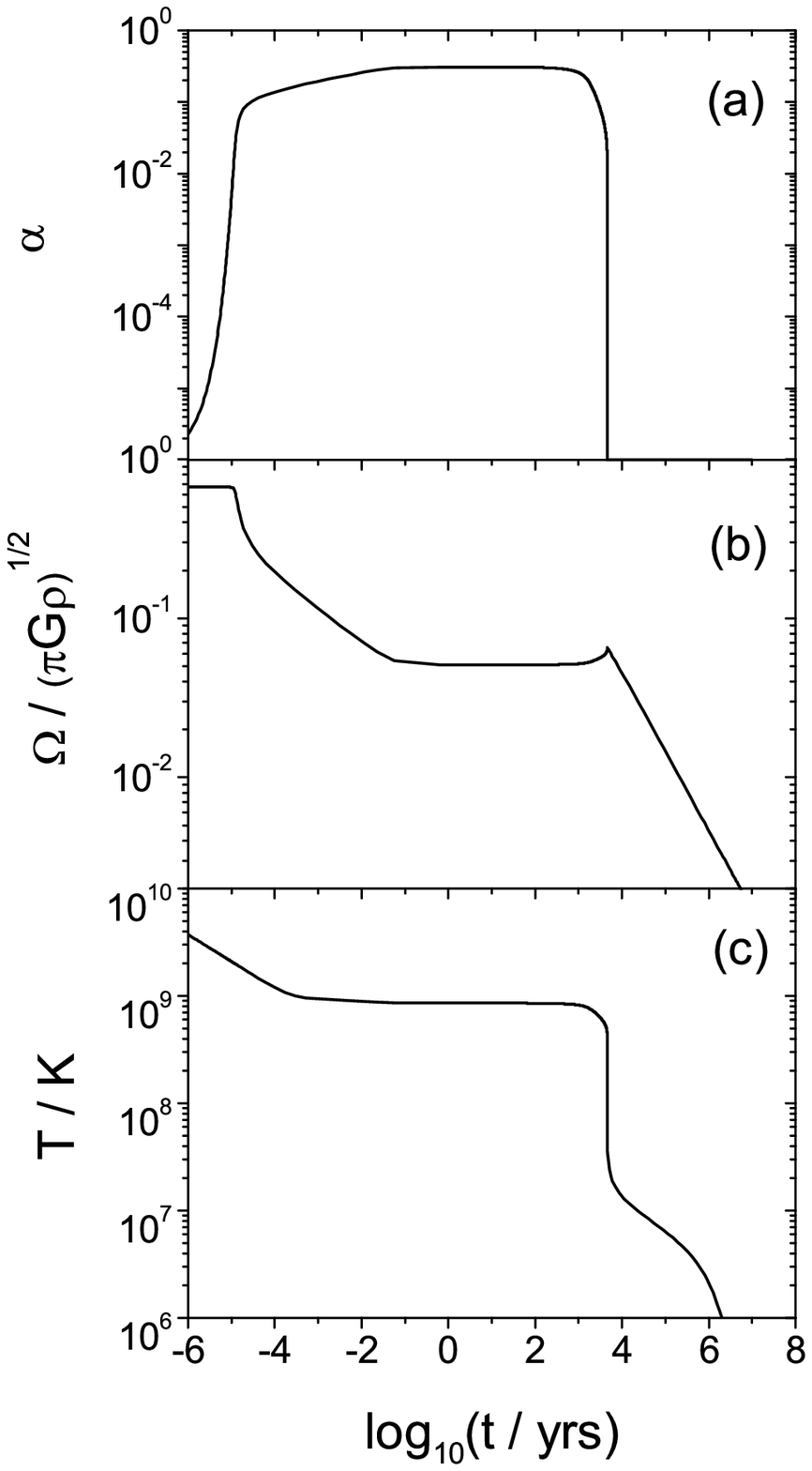}
   \caption{Evolution curves of $\alpha$, $\Omega$, and $T$ of a
    $1.4M_{\odot}$ NS with magnetic field $B=10^{12}$ G and $K=1000$.}
   \label{Fig:f1}
   \end{figure}

 \begin{figure}
   \centering
   \includegraphics[width=0.7\textwidth]{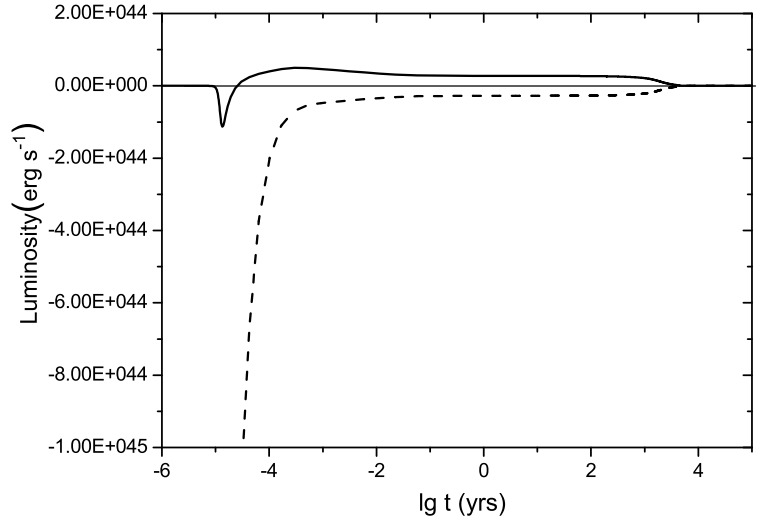}
   \caption{Evolution curves of $(H_{v}-\Delta L_{\nu})$ due to viscous dissipation (thick solid line) and ($-L_{\nu} -L_{\gamma}$) (dashed line) with the same parameters as Fig.1. The thin solid one is the zero luminosity line.}
   \label{Fig:f2}
   \end{figure}

\begin{figure}
   \centering
   \includegraphics[width=0.7\textwidth]{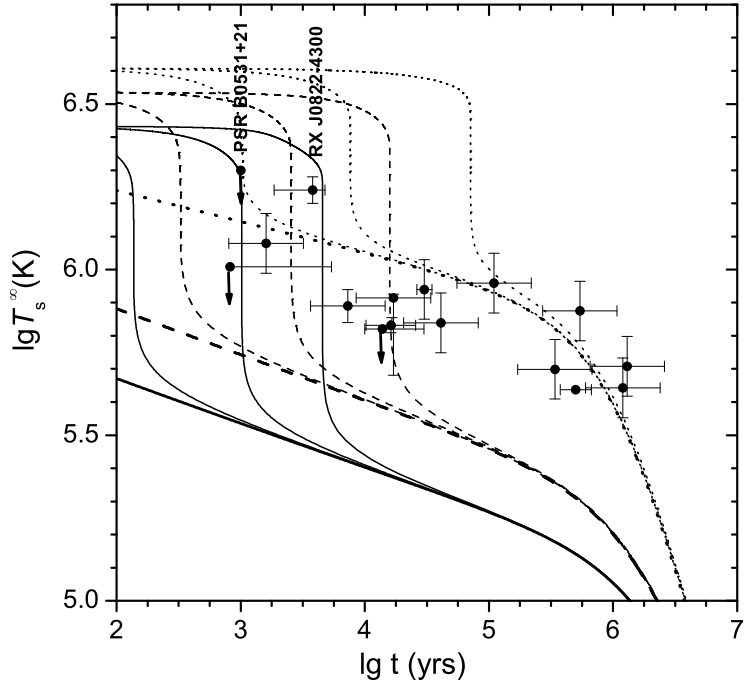}
   \caption{Observational data on the surface temperatures of NSs (Yakovlev et al. 2008) compared with
   theoretical cooling curves. The dot, dashed and solid curves correspond to $M=1.3M_{\odot}$, $M=1.365M_{\odot}$
   and $M=1.4M_{\odot}$, respectively. The thick lines are calculated without the \textit{r}-mode dissipation effect, and the  thin lines refer to $K=10$, $100$, $1000$ form left to right. For all curves, $B=10^{12}$G. }
   \label{Fig:f3}
   \end{figure}
\end{document}